\documentclass[11pt]{article}

\usepackage{graphicx,epsfig}
\usepackage{multicol}
\usepackage{amsmath,amssymb,cite,color,hyperref}
\newcommand{\be}{\begin{equation}}
\newcommand{\ee}{\end{equation}}
\newcommand{\bea}{\begin{eqnarray}}
\newcommand{\eea}{\end{eqnarray}}
\newcommand{\bi}{\begin{itemize}}
\newcommand{\ei}{\end{itemize}}
\newcommand{\ben}{\begin{enumerate}}
\newcommand{\een}{\end{enumerate}}

\newcommand{\lc}{\left[}
\newcommand{\rc}{\right]}
\newcommand{\lp}{\left(}
\newcommand{\rp}{\right)}

\def\frac#1#2{{{#1}\over {#2}}}
\def\gsim{\mathrel{\rlap{\lower4pt\hbox{\hskip1pt$\sim$}}
    \raise1pt\hbox{$>$}}}         
\def\lsim{\mathrel{\rlap{\lower4pt\hbox{\hskip1pt$\sim$}}
    \raise1pt\hbox{$<$}}}         

\newcommand{\draft}[1]{}

\definecolor{grey}{rgb}{0.5,0.5,0.5}

\begin{document}
\begin{flushright}
Edinburgh 2011/08\\
IFUM-973-FT\\
FR-PHENO-2010-003\\
RWTH TTK-11-04\\
\end{flushright}
\begin{center}

{\Large \bf On the Impact of NMC Data\\ on NLO and NNLO Parton
Distributions \\\medskip
and Higgs Production at the Tevatron and the LHC}\vspace{0.8cm}

{\bf  The NNPDF Collaboration:}\\
Richard~D.~Ball$^{1}$, Valerio~Bertone$^3$, Francesco~Cerutti$^4$,
 Luigi~Del~Debbio$^1$,\\ Stefano~Forte$^2$, Alberto~Guffanti$^3$, 
Jos\'e~I.~Latorre$^4$, Juan~Rojo$^2$ and Maria~Ubiali$^{5}$.

\vspace{1.cm}
{\it ~$^1$ School of Physics and Astronomy, University of Edinburgh,\\
JCMB, KB, Mayfield Rd, Edinburgh EH9 3JZ, Scotland\\
~$^2$ Dipartimento di Fisica, Universit\`a di Milano and
INFN, Sezione di Milano,\\ Via Celoria 16, I-20133 Milano, Italy\\
~$^3$  Physikalisches Institut, Albert-Ludwigs-Universit\"at Freiburg,\\ 
Hermann-Herder-Stra\ss e 3, D-79104 Freiburg i. B., Germany  \\
~$^4$ Departament d'Estructura i Constituents de la Mat\`eria, 
Universitat de Barcelona,\\ Diagonal 647, E-08028 Barcelona, Spain\\
~$^5$ Institut f\"ur Theoretische Teilchenphysik und Kosmologie, RWTH Aachen University,\\ 
D-52056 Aachen, Germany\\}
\end{center}

\vspace{0.8cm}

\begin{center}
{\bf \large Abstract:}
\end{center}

We discuss the impact of the treatment of 
NMC structure function data on parton distributions in the context of
the NNPDF2.1 global PDF determination at NLO and NNLO. 
We show that the way these data are treated, and even their complete
removal, has  no effect  on parton distributions at NLO, and at NNLO has
an effect which is below one sigma. In particular, the Higgs production
cross-section in the gluon fusion channel is very stable.

\clearpage

Fixed target deep--inelastic scattering data
provide important constraints on parton distributions (PDFs) and are
routinely included in PDF determinations.
It has been recently suggested~\cite{Alekhin:2011ey} 
that the results of current PDF determinations
depend strongly on the treatment of the fixed target 
DIS data obtained by the NMC
collaboration~\cite{Arneodo:1996qe,Arneodo:1996kd}: in particular,
according to whether data for cross-sections or structure functions are used in the fit.
The substantial changes in the gluon distribution and $\alpha_s\lp M_Z\rp$ found in
Ref.~\cite{Alekhin:2011ey} lead to a large shift in the Higgs production cross-section,
which would, if correct, have very significant implications for Higgs searches 
at the Tevatron and LHC. This claim has generated an ongoing 
discussion on the
adequacy of current Higgs mass
limits~\cite{Baglio:2011hc,Thorne:2011kq}; besides its interest in
this context, the issue is relevant for the understanding of the
comparative merits of PDF determinations based on a wider dataset
(which contain more information but might be less consistent) and
those based on a more limited but more consistent set of data.

In this note we examine this issue within the context of the
NNPDF2.1 NLO~\cite{Ball:2011mu}  and NNLO~\cite{Ball:2011uy} PDF
determinations. In contrast to the ABKM~\cite{Alekhin:2009ni}
determination, on which the results of Ref.~\cite{Alekhin:2011ey} are
based, NNPDF2.1 depends on a rather broader dataset, and uses the especially
flexible NNPDF methodology (for a review see
e.g. Ref.~\cite{Forte:2010dt}), making it less vulnerable to
parametrization bias. Related results (consistent with our
findings) have been presented recently in the context of the
MSTW~\cite{Thorne:2011kq} and CTEQ~\cite{HustonPDF4LHC} PDF determinations.

\begin{figure}[t]
\begin{center}
\epsfig{file=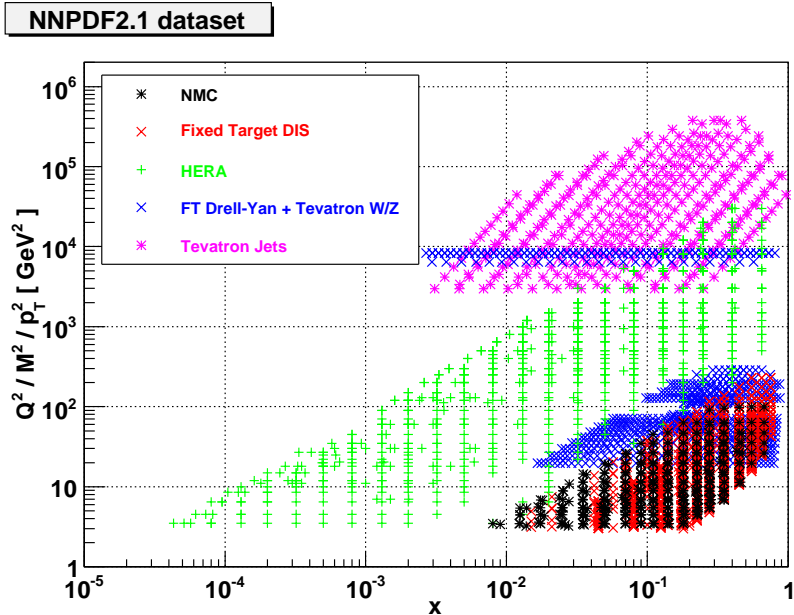,width=0.78\textwidth}
\caption{\small 
Kinematic coverage of the NMC data, compared
to those of the other datasets in the NNPDF2.1 global analysis:
the non-NMC fixed target DIS data, the HERA collider data,
the fixed target Drell-Yan and Tevatron weak vector boson
production data and the Tevatron inclusive jet data.
\label{fig:kinplot} }
\end{center}
\end{figure}

The kinematic coverage of the NMC data is compared
in Fig.~\ref{fig:kinplot} to that of other datasets used to determine
NNPDF2.1 PDFs: 
the other  fixed target DIS data, the HERA collider data,
the fixed target Drell-Yan and Tevatron weak vector boson
production data and the Tevatron inclusive jet data. We will now
consider variants of NNPDF2.1 in which the NMC data are treated in
different ways. In all other respects, we adopt the default
settings of NNPDF2.1 as discussed in Refs.~\cite{Ball:2011mu,Ball:2011uy}. In
particular,  we take a fixed value for the
strong coupling in both the NLO and NNLO fits, 
$\alpha_s\lp M_Z\rp=0.119$, close
to the PDG average~\cite{Bethke:2009jm}; sets with variable 
$\alpha_s\lp M_Z\rp$ are also 
available~\cite{Demartin:2010er,Ball:2010de,Ball:2011mu,Ball:2011uy}, 
from which 
combined PDF+$\alpha_s$ uncertainties can be
computed~\cite{Alekhin:2011sk,Dittmaier:2011ti}. 

The  NMC collaboration has measured 
the neutral current deep-inelastic muon-nucleon 
cross-section
\be
\frac{d^2\sigma^{\rm NC}}{dxdQ^2} (x,y,Q^2)=\frac{2\pi \alpha^2}{ x Q^4}
\lc
Y_+ F_2^{NC}(x,Q^2) \mp Y_- x F_3^{NC}(x,Q^2)-y^2 F_L^{NC}(x,Q^2)\rc\,,
\label{eq:ncxsect}
\ee
where $Y_{\pm}=1\pm (1-y)^2$. For NMC $Q^2\ll M_W^2$
so the parity--violating structure function
$xF_3$ can be neglected and only the electromagnetic
components of $F_2$ and $F_L$ are relevant.
It is convenient to define a 
 reduced cross-section 
\begin{equation}
  \label{eq:rednc}
  \widetilde{\sigma}^{\rm NC}(x,y,Q^2)=\lc
  \frac{2\pi \alpha^2}{ x Q^4} Y_+\rc^{-1}\frac{d^2\sigma^{\rm NC}}{dxdQ^2}(x,y,Q^2)=F_2^{\rm NC}(x,Q^2)
\Big(2 -2y +\frac{y^2}{1+R(x,Q^2)}\Big)\,,
\end{equation}
where
\be
\label{eq:r}
R\lp x,Q^2\rp=F_L(x,Q^2)/\lp F_2(x,Q^2)-F_L(x,Q^2) \rp \,.
\ee 
Equation~(\ref{eq:rednc}) was used by the NMC collaboration~\cite{Arneodo:1996qe,Arneodo:1996kd} to extract
$F_2^{\rm p}$ from the measured cross-section Eq.~(\ref{eq:ncxsect}), using
for $x\le 0.12$ a determination of $R(x,Q^2)$
from their own data, and  for $x\ge 0.12$ a parametrization $R_{1990}$
of $R$~\cite{Whitlow:1991uw} obtained from a global  fit to SLAC
structure function data~\cite{Whitlow:1990gk}.

In all NLO NNPDF parton determinations~\cite{DelDebbio:2007ee,Ball:2008by,Ball:2009mk,Ball:2010de,Ball:2011mu}
the NMC structure function data was used, both for the proton
structure function $F_2^{\rm p}$ and the ratio of deuteron
to proton structure functions, $F_2^{\rm d}/F_2^{\rm p}$. It may be reasonably argued
however that data for the reduced cross-section, which is closer to what is
measured experimentally, should be used instead. 
Note that the distinction is only relevant for the $F_2^{\rm p}$ data:
since the isotriplet component of $F_L\lp x,Q^2\rp$ is very small, 
$R^p(x,Q^2)\approx R^d(x,Q^2)$, so 
\begin{equation}
  \label{eq:sf-ratio}
  \frac{\widetilde{\sigma}^{\rm d}(x,y,Q^2)}{
\widetilde{\sigma}^{\rm p}(x,y,Q^2)}\approx \frac{F_2^{\rm d}(x,Q^2)}{F_2^{\rm p}(x,Q^2)}
 \,.
\end{equation}

\begin{figure}[t]
\begin{center}
\epsfig{file=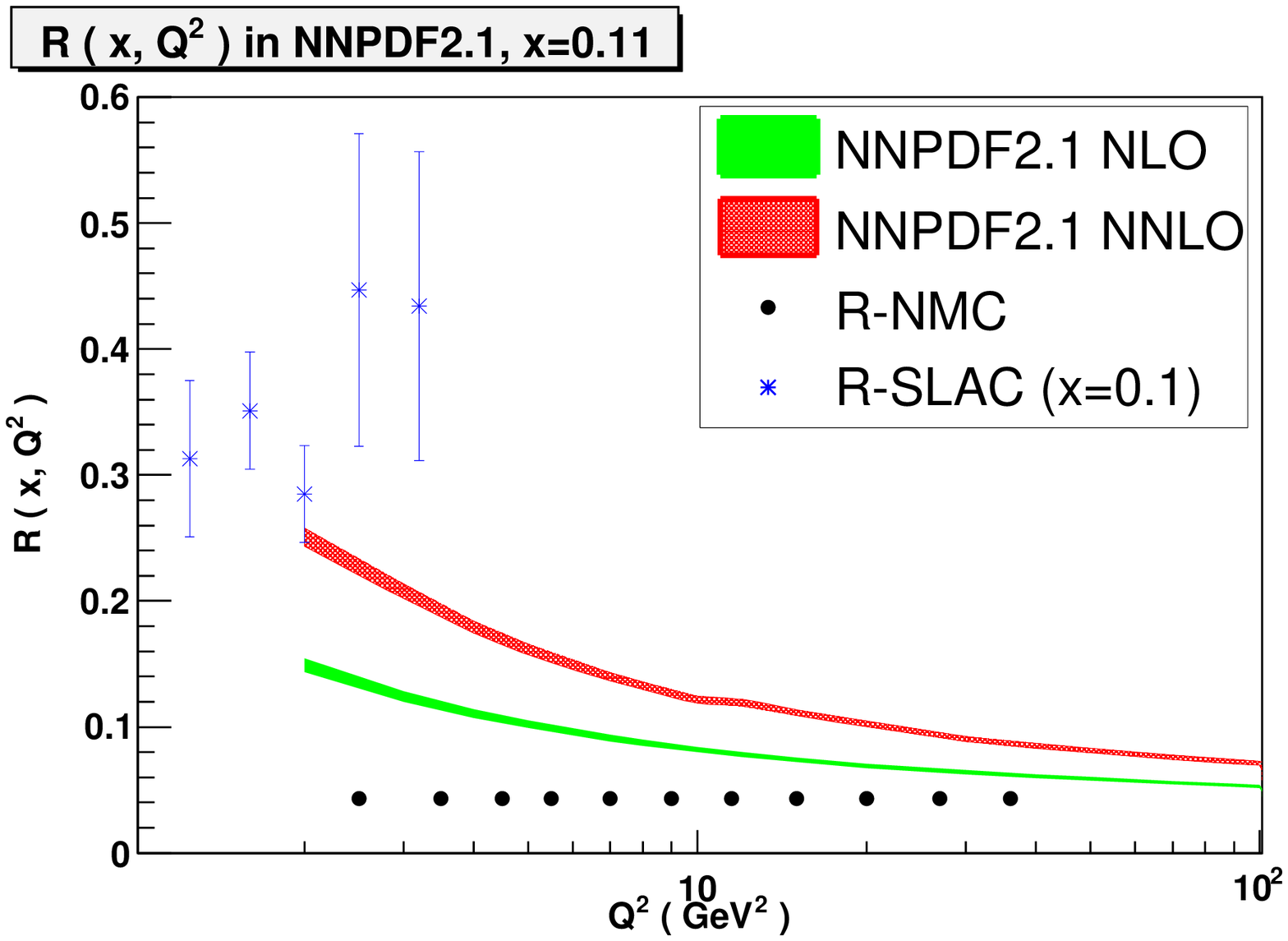,width=0.49\textwidth}
\epsfig{file=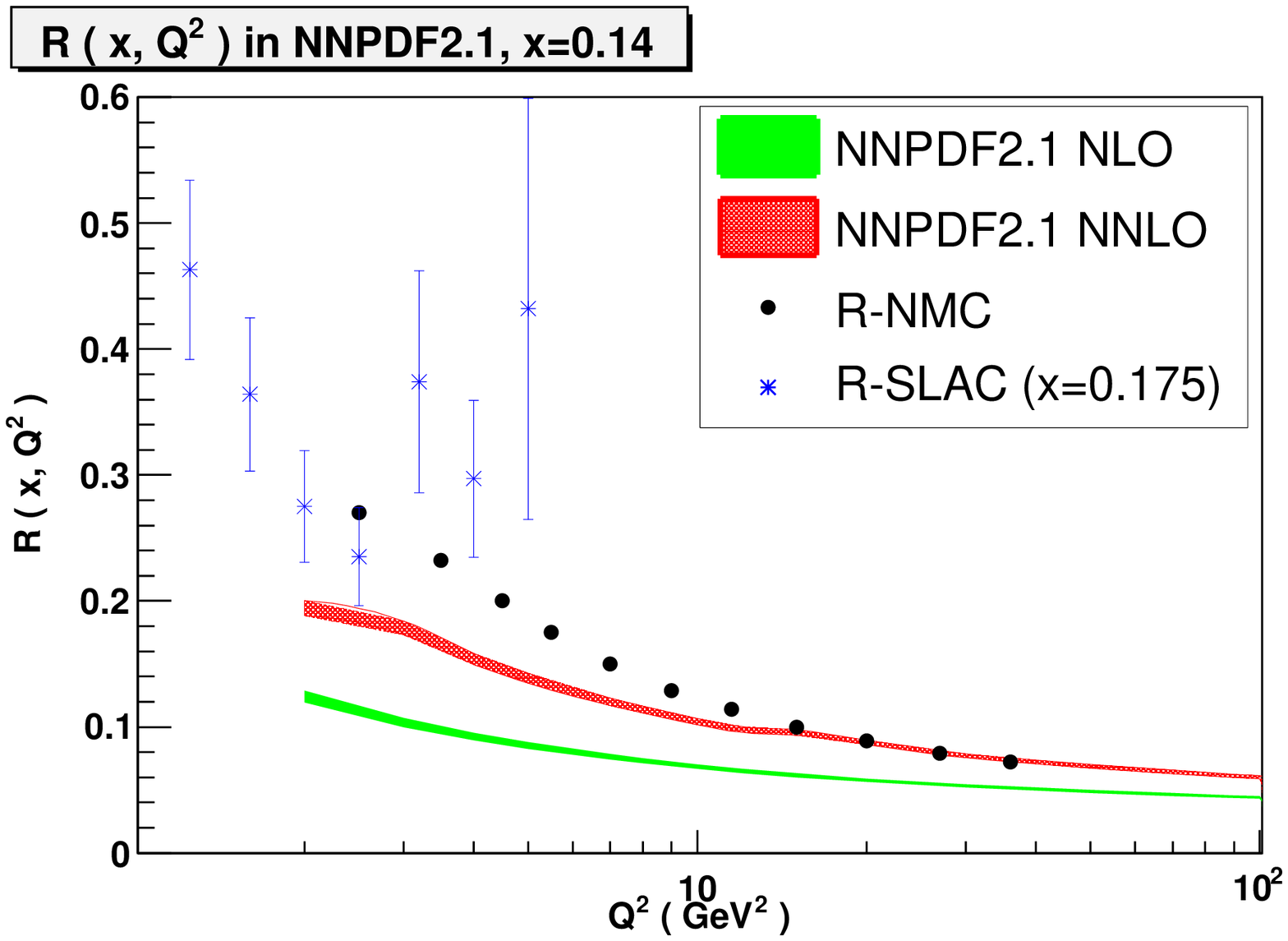,width=0.49\textwidth}
\caption{\small 
The NNPDF2.1 NLO and NNLO results
for $R(x,Q^2)$ Eq.~(\ref{eq:r}) at $x=0.11$ (left) and
$x=0.14$ (right), compared to the values of $R$ used by  NMC in
Ref.~\cite{Arneodo:1996qe}, and the SLAC data of  Ref.~\cite{Whitlow:1990gk} on
which the parametrization~\cite{Whitlow:1991uw} used by NMC for
$x>0.12$ is based.
\label{fig:rnmc} }
\end{center}
\end{figure}
In Fig.~\ref{fig:rnmc} (to be compared to Fig.~(1) of
Ref.~\cite{Alekhin:2009ni}) the form of $R(x,Q^2)$ used by NMC (shown as black
dots) in both regions is compared to the prediction obtained using 
NNPDF2.1 NLO and NNLO PDF sets. The parametrization $R_{1990}$ does
not come with an uncertainty, however the typical size of the
uncertainty on it can be inferred by comparing it to the data of
Ref.~\cite{Whitlow:1990gk} on which it is based, also shown in
Fig.~\ref{fig:rnmc}. Note that the SLAC data are concentrated at low
$Q^2$, hence in most of the NMC kinematic region this 
parametrization is  an extrapolation and thus subject to very large uncertainties.
 It is clear from Fig.~\ref{fig:rnmc} that (as emphasized in
Ref.~\cite{Alekhin:2009ni}) the $R$ values used by NMC at low $x$ do
not agree well with the prediction from the use of modern PDF sets,
while instead the parametrization $R_{1990}$ is in good agreement with
the NNPDF prediction within the large uncertainty of the data on which
it is based, especially if NNLO theory is used. Thus the use of NMC
cross-sections 
instead of structure functions (which rely on these partly inadequate
assumptions on $R$) does indeed appear to be in principle more advisable.

\begin{figure}[t]
\begin{center}
\epsfig{file=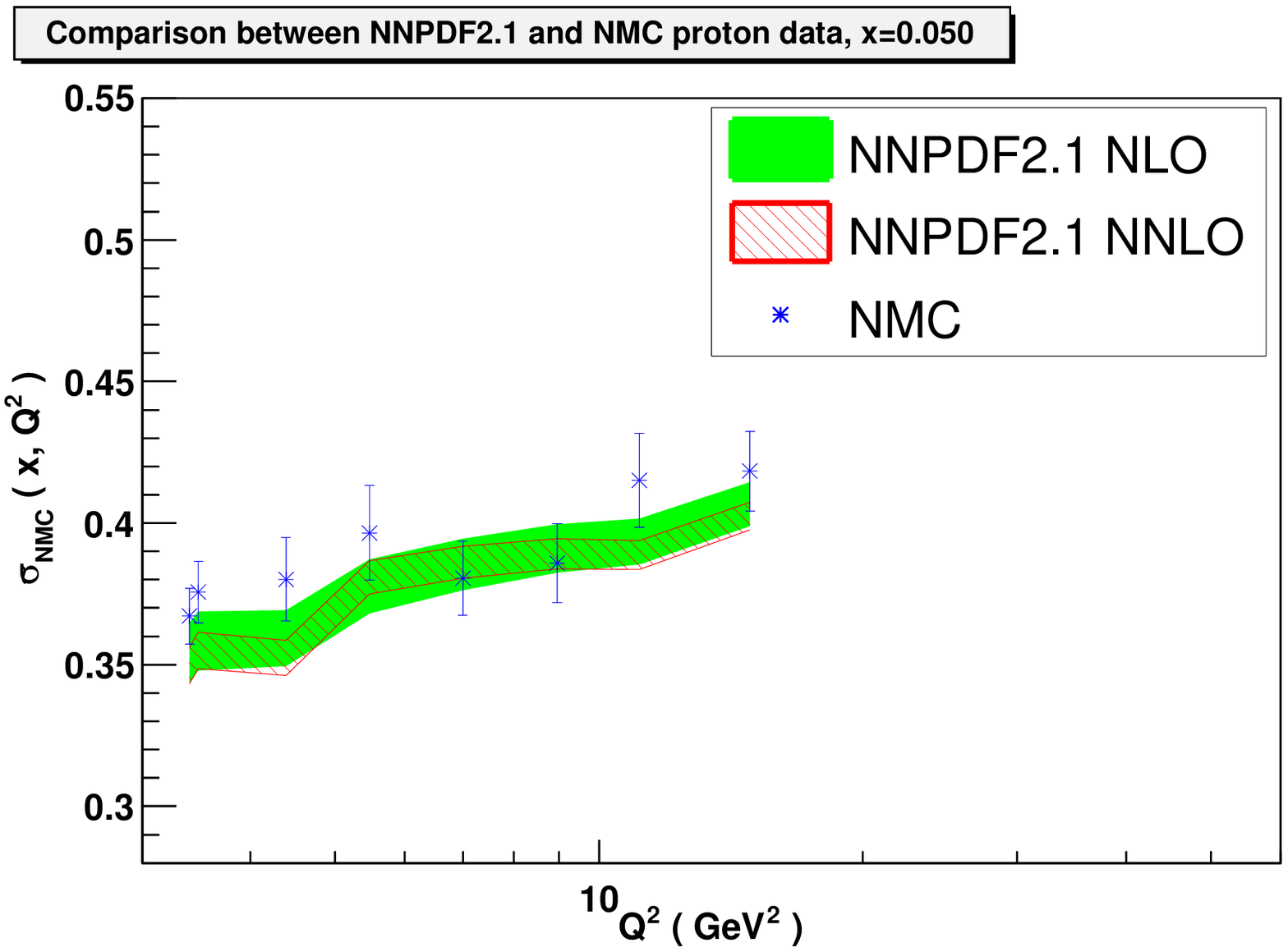,width=0.32\textwidth}
\epsfig{file=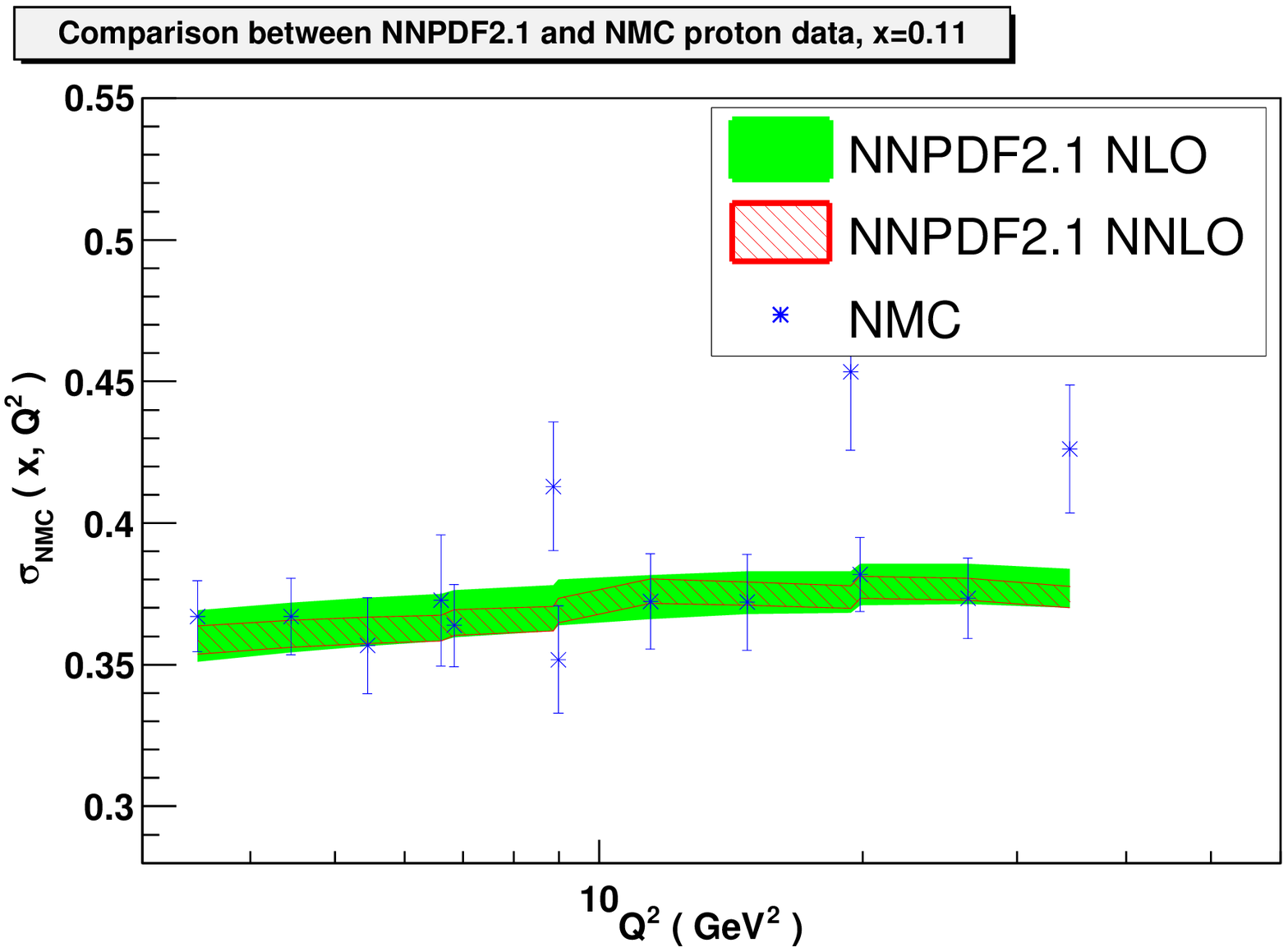,width=0.32\textwidth}
\epsfig{file=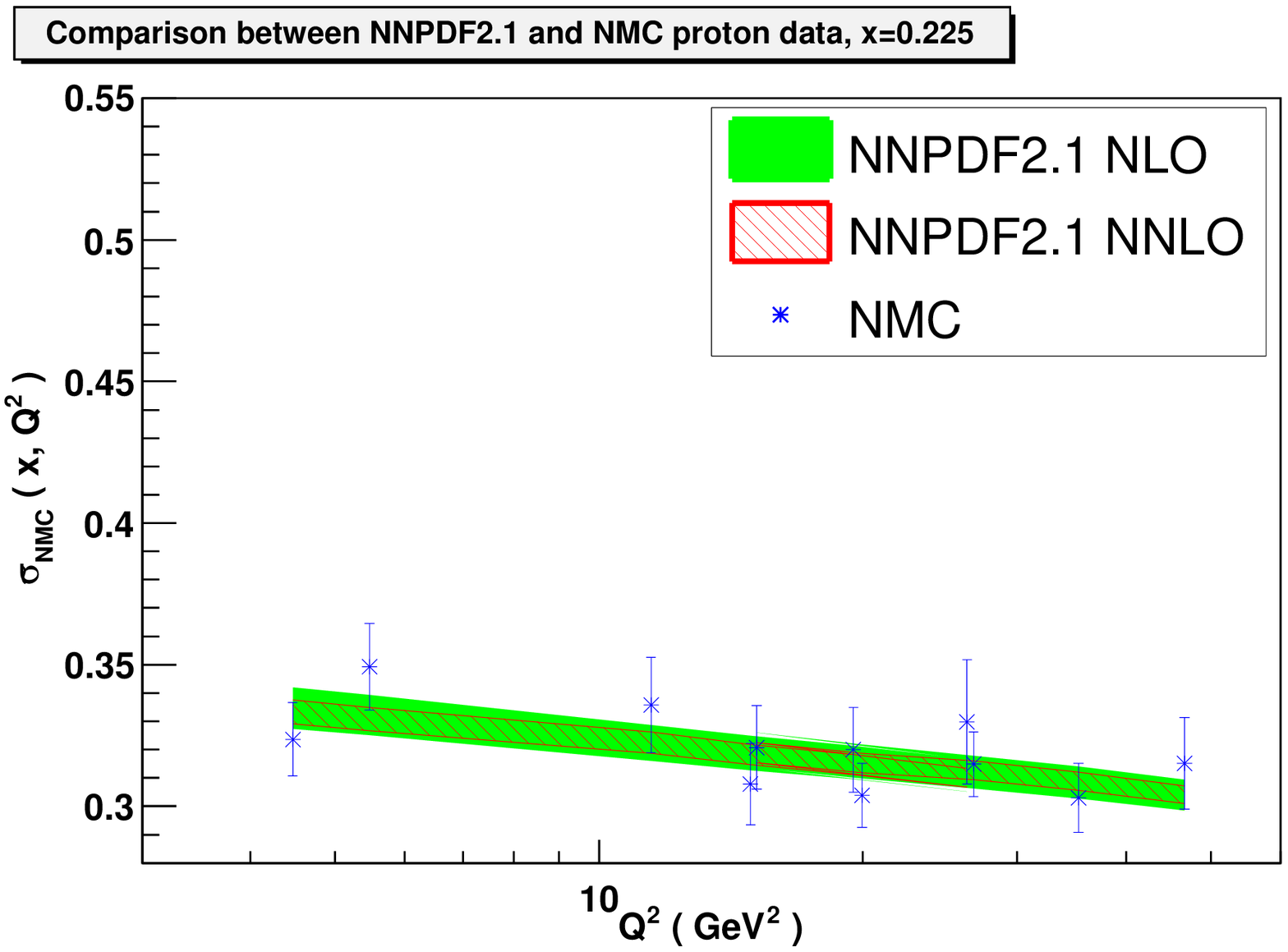,width=0.32\textwidth}
\caption{\small 
The NNPDF2.1 NLO and NNLO predictions
for the NMC reduced cross-sections for $x=0.05$ (left),
$x=0.11$ (center) and
$x=0.225$ (right ), compared to the NMC data~\cite{Arneodo:1996qe}.
\label{fig:f2nmc} }
\end{center}
\end{figure}

However, it is unclear whether in practice the effect of this
replacement may be significant, especially in view of the fact that  
the NMC data are known to  have
internal consistency problems, as shown long ago in
Ref.~\cite{Forte:2002fg}. To illustrate this, in Fig.~\ref{fig:f2nmc}
the NMC reduced cross-section data are compared to NLO and NNLO
predictions obtained using the corresponding NNPDF2.1 PDF sets. It is
clear that the data show larger point-by-point fluctuations than one
would expect from their nominal uncertainties, thereby suggesting
that the effect of the treatment of the relatively small
$R$--dependent correction might be moderate on the scale of these fluctuations.

In order to settle the issue quantitatively, 
we construct  and compare, both at NLO and at NNLO, three PDF sets: one in which
NMC data for the 
proton structure function $F_2^{\rm p}$ are used, one in which
data for the proton reduced cross-section are used (supplemented, in
both cases, by data for the
ratio $F_2^{\rm d}/F_2^{\rm p}$), 
 and one in which NMC data (both for proton and the deuteron/proton
ratio)
are removed altogether from the global dataset.
In all cases sets of $N_{\rm rep}=100$ replicas have been
produced. Note that the published (default) NNPDF2.1 
sets~\cite{Ball:2011mu,Ball:2011uy} use NMC structure
functions at NLO and NMC cross-sections at NNLO.

\begin{table}[t]
\begin{center}
\footnotesize
\begin{tabular}{|c||c|c|c||c|c|c|}
\hline
 & \multicolumn{3}{|c|}{ NNPDF2.1 NLO} & \multicolumn{3}{|c|}{ NNPDF2.1 NNLO}  \\
\hline
&            str. fctn.  & xsec. & noNMC 
&   str. fctn.  & xsec.  &  noNMC \\
\hline
\hline
Total      &  1.16   &      1.14      &     1.09 & 1.16 & 1.16 & 1.12 \\
\hline
NMC-pd    &   0.97    &     0.98       &   -  & 0.93& 0.93 & - \\
{\it NMCp}      &  {\it 1.73}     &  {\it   1.67}    &  -    &   {\it   1.69}   &
{\it   1.63}   & - \\
SLAC      &   1.27    &     1.27          &   1.28   & 1.05 &1.01  & 1.00 \\
BCDMS     &    1.24   &      1.23           &  1.18  & 1.29&  1.32 &  1.27 \\
HERAI-AV  &   1.07    &     1.05           &    1.07  & 1.12 & 1.10 & 1.08 \\
CHORUS    &   1.15    &     1.11            &   1.07  & 1.12 & 1.12 & 1.12 \\
FLH108    &   1.37    &     1.34            &  1.38 & 1.27 & 1.26 &  1.29 \\
NTVDMN    &   0.47    &     0.51        &   0.42  & 0.50 & 0.49 & 0.50 \\
ZEUS-H2   &   1.29    &     1.23         &  1.24   & 1.32 & 1.31 & 1.30\\
ZEUSF2C   &   0.78    &     0.74         &   0.72 & 0.88 & 0.88 & 0.89 \\
H1F2C     &   1.51    &     1.48          &   1.49  & 1.47 & 1.56 & 1.52\\
DYE605    &   0.85    &     0.93            &  0.88  & 0.81& 0.81 &  0.81 \\
DYE866    &   1.27    &     1.40             & 1.34   & 1.31 & 1.32 & 1.34 \\
CDFWASY   &   1.85    &     1.87             &  1.60  &  1.55 & 1.65 & 1.41 \\
CDFZRAP   &   1.62    &     1.76              &   1.64  &  2.16 & 2.12 & 2.18\\
D0ZRAP    &   0.60    &     0.57              &   0.56  & 0.67 & 0.67 & 0.67\\
CDFR2KT   &   0.97    &     0.73           &    0.81   & 0.79 & 0.74 & 0.80\\
D0R2CON   &   0.84    &     0.90               &  0.96  & 0.84 & 0.82  &  0.84  \\
\hline
\end{tabular}
\end{center}
\caption{\small The $\chi^2$ of NNPDF2.1 PDFs at NLO and NNLO when NMC
  data are included as structure functions, reduced cross-sections, or
  not included.
\label{tab:chi2tab}}
\end{table}

In Table~\ref{tab:chi2tab}
we compare the $\chi^2$ values obtained in these three fits, both for
the global fit and individual experiments.
The quality of the global fit is unchanged at NNLO and improves
slightly at NLO when the structure function data are replaced by cross-section data, and in both cases the quality of the fit to NMC data
improves slightly, with the quality of the fit to other data
unchanged. This suggests that  the use of cross-section data is
indeed somewhat more consistent for NMC, but also that this has little or no
effect on other experiments. 
When the NMC
data is removed altogether, the global fit quality 
improves, due to the fact that $\chi^2_{\rm NMC}$ is rather poor in
view of the aforementioned inconsistencies, regardless of
how NMC data are treated. In
particular, the fit to the BCDMS data, which measure the same
structure functions
as NMC in a partly overlapping region, improves somewhat when the NMC
data are removed. 

\begin{figure}[t]
\begin{center}
\epsfig{file=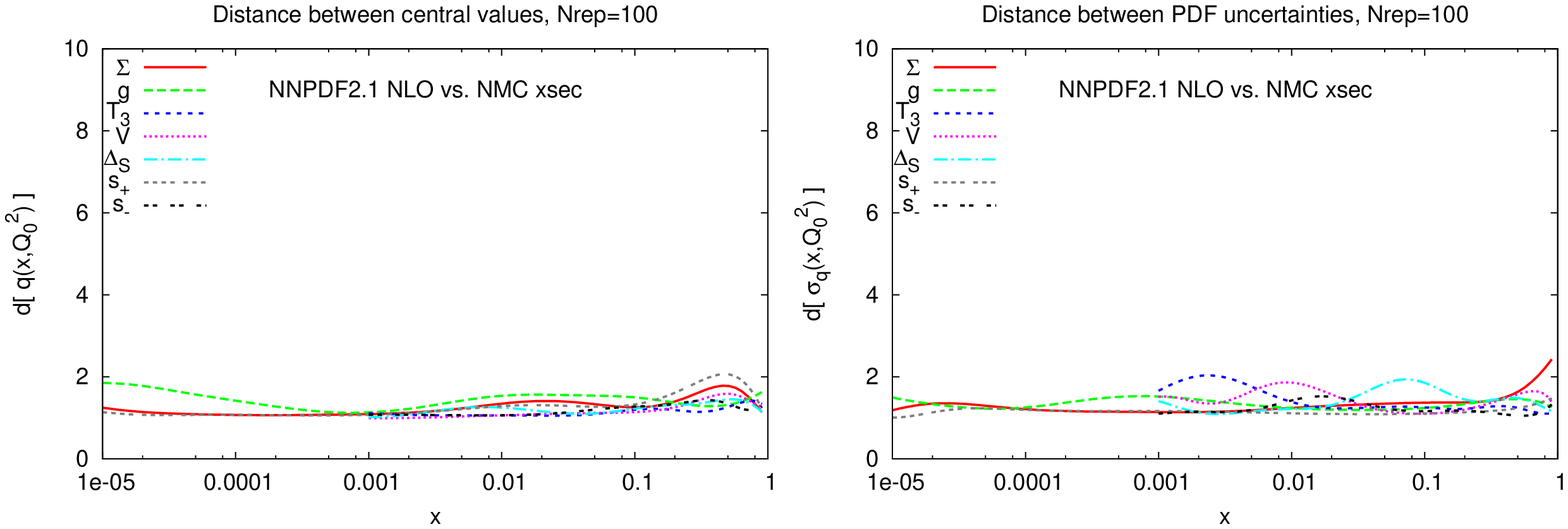,width=\textwidth}
\epsfig{file=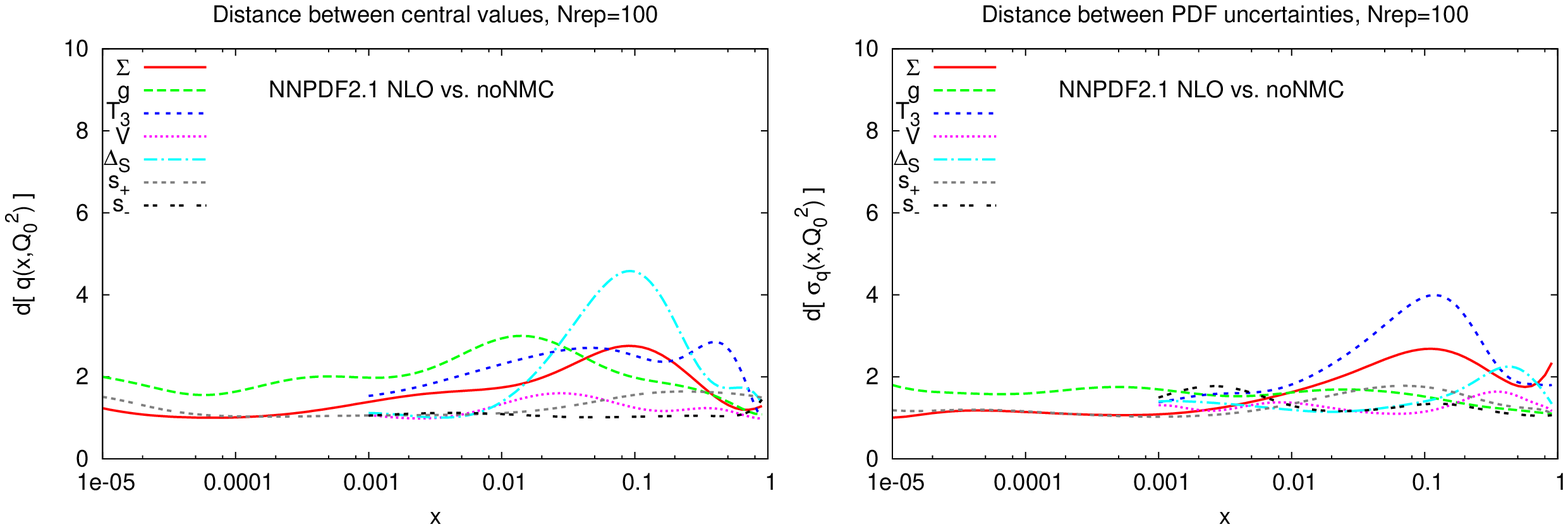,width=\textwidth}
\caption{\small 
Distances (defined as in Ref.~\cite{Ball:2010de}) between NLO PDF sets
with NMC structure functions and NMC cross-sections (top) and PDF sets
with NMC structure functions and PDF sets without NMC data (bottom). All
distances have been computed using
sets
of $N_{\rm rep}=100$ replicas. 
\label{fig:distances} }
\end{center}
\end{figure}
\begin{figure}[t]
\begin{center}
\epsfig{file=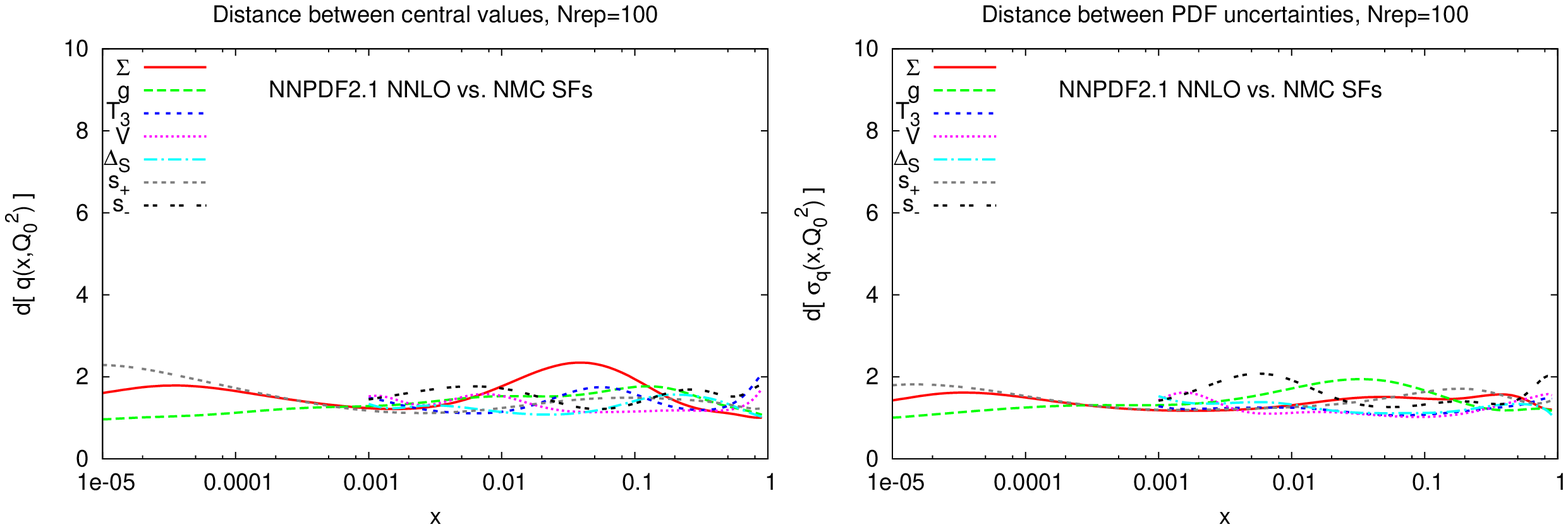,width=\textwidth}
\epsfig{file=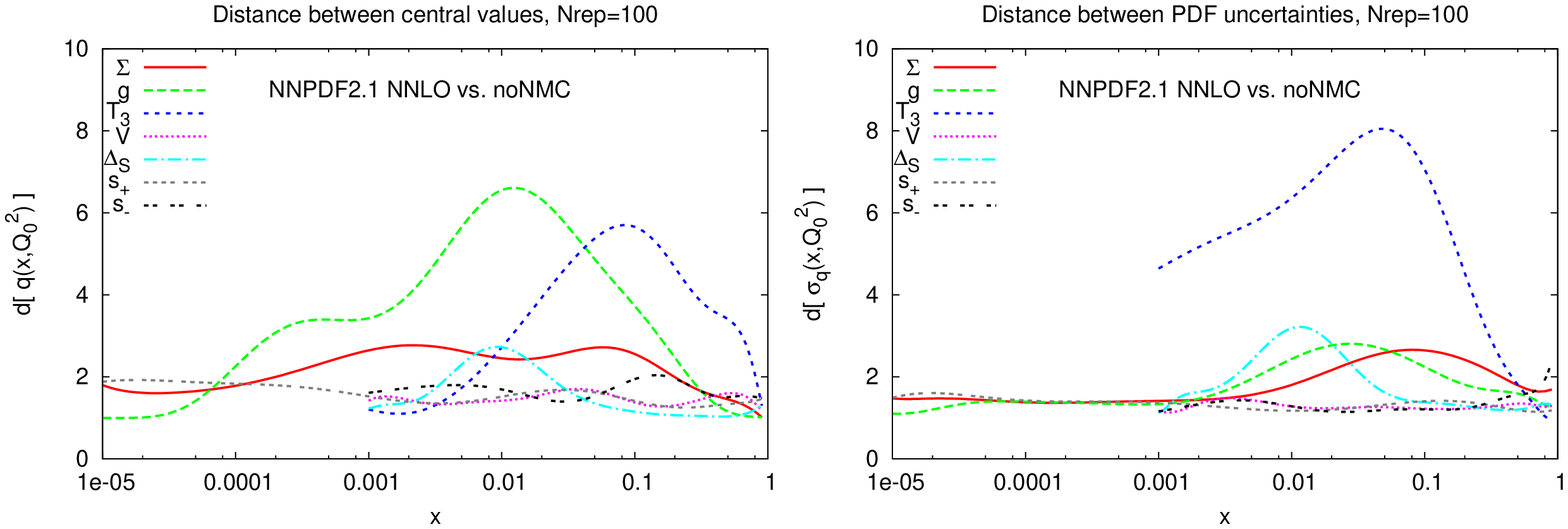,width=\textwidth}
\caption{\small 
Same as Fig.~\ref{fig:distances} but at NNLO. Note that in this case
the PDF set with NMC cross-section data (NNLO default) is used for the
comparison with the PDFs with no NMC data.
\label{fig:distances-nnlo} }
\end{center}
\end{figure}

We now compare the PDFs obtained in the various cases.
In Fig.~\ref{fig:distances} (NLO)
and Fig.~\ref{fig:distances-nnlo} (NNLO) we show the 
distances  (as defined in Ref.~\cite{Ball:2010de}), computed both for
central values and uncertainties,
between  PDFs in
the sets with NMC cross-section vs. structure function data, and a set
without NMC data vs. the
the default NNPDF2.1 set. 
Recall that $d\sim 1$ corresponds
to statistically indistinguishable results, while, for sets of 100
replicas, $d\sim 7$ corresponds to a shift by one sigma
(i.e. results are statistically distinguishable, but compatible within
uncertainties).

These plots show that at NLO the replacement of structure functions with
cross-sections is at the level of statistical fluctuations. At NNLO a
small, statistically significant, 
shift in central values and uncertainties at the level of at most
a third of a sigma but mostly lower  is seen in some PDFs
(specifically the quark singlet and isospin triplet and the gluon).
The effect of removing NMC data altogether at NLO is again almost
indistinguishable from a statistical fluctuation  with the possible
exception of the quark singlet for $0.02\lsim x\lsim 0.5$ which shows
a shift by little more than a quarter of standard deviation (though
this could be a statistical fluctuation due to the size of the
replica sample). At NNLO instead the effect of removing the NMC data
altogether is clearly statistically significant on the isospin triplet
and gluon, corresponding to a shift in central values 
at the level of almost one sigma for the gluon and more
than half sigma for the isospin triplet. A one sigma change of the
triplet uncertainty is also observed.

\begin{figure}[t]
\begin{center}
\epsfig{file=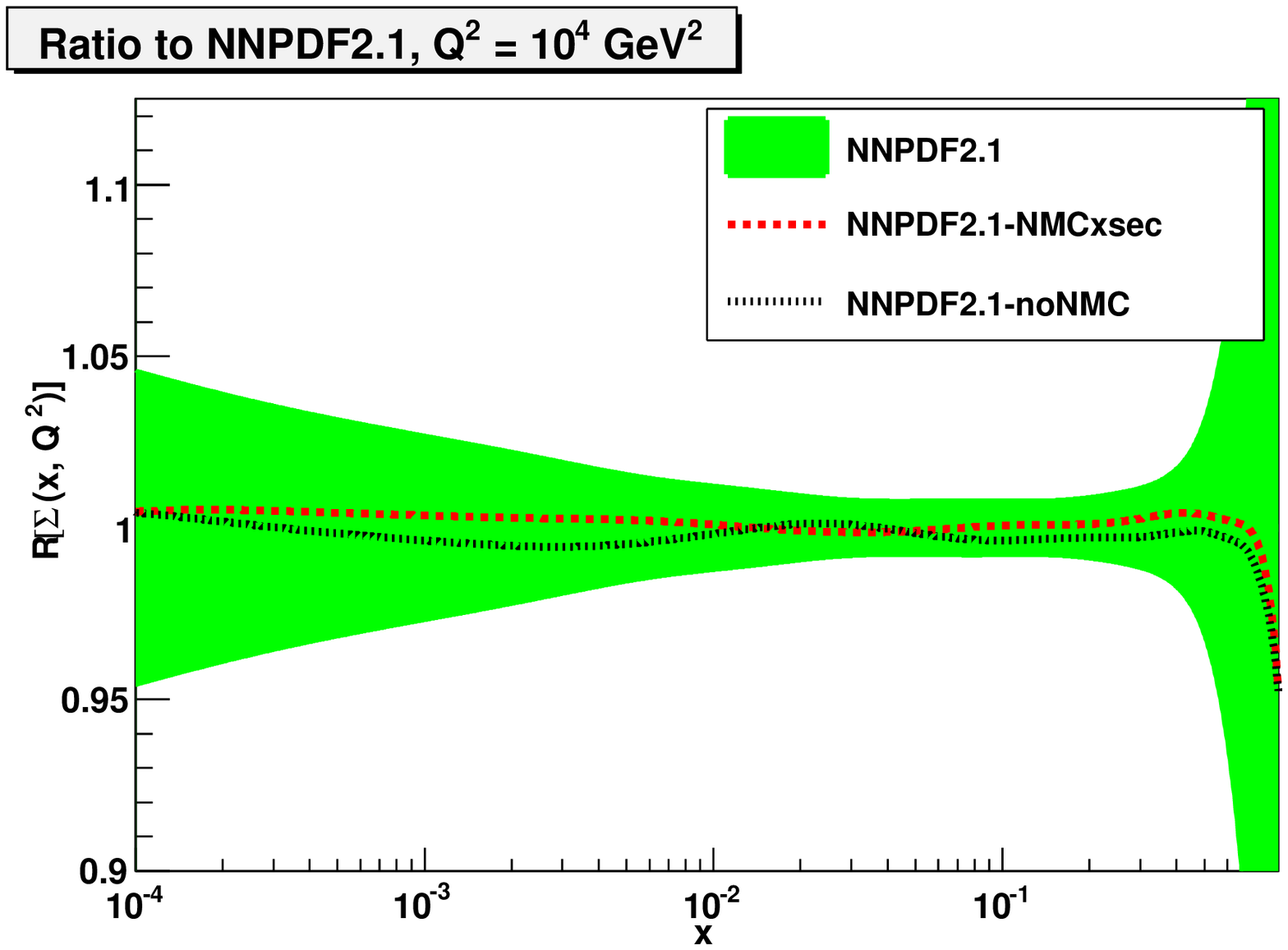,width=0.48\textwidth}
\epsfig{file=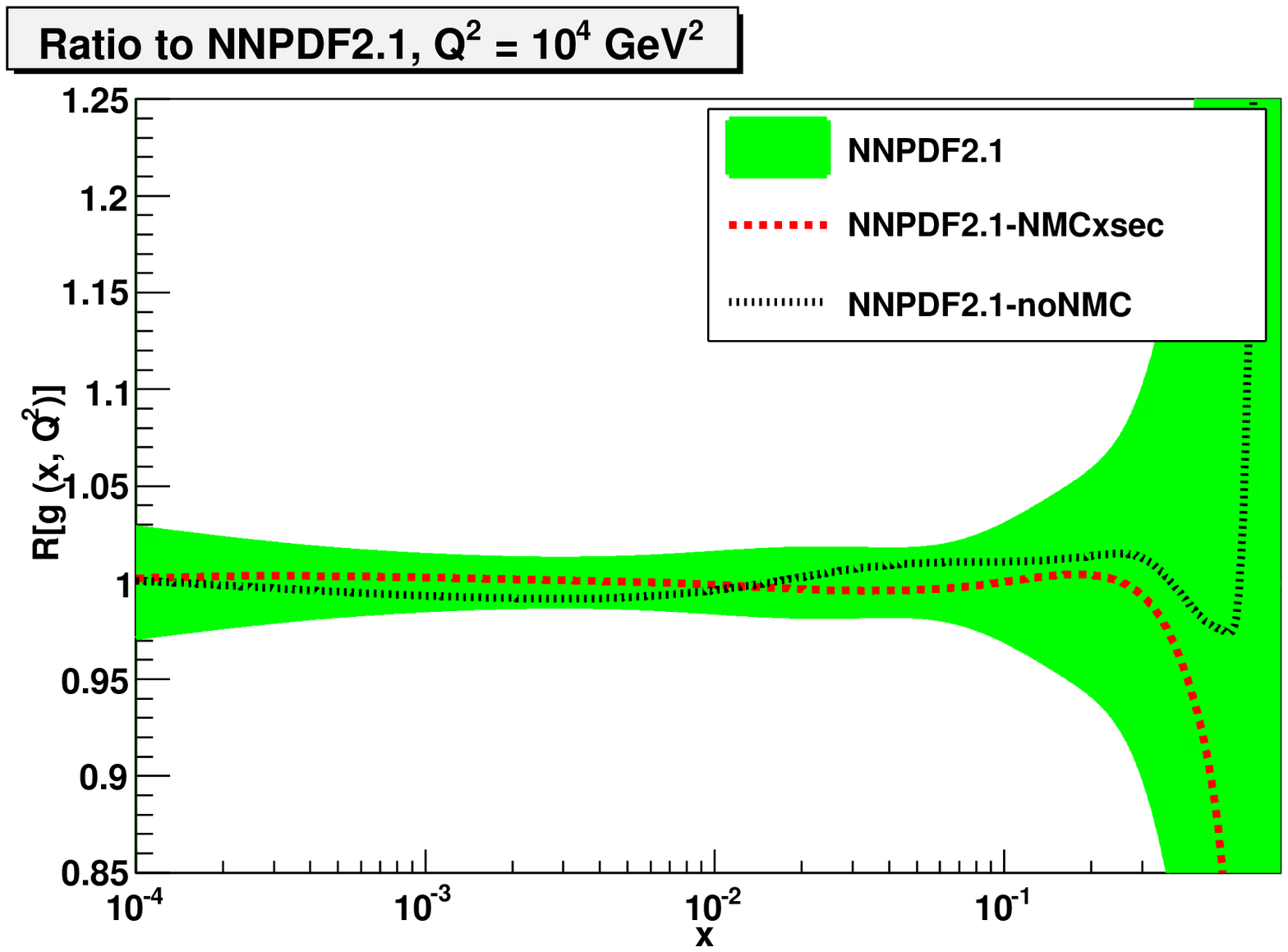,width=0.48\textwidth} \\
\epsfig{file=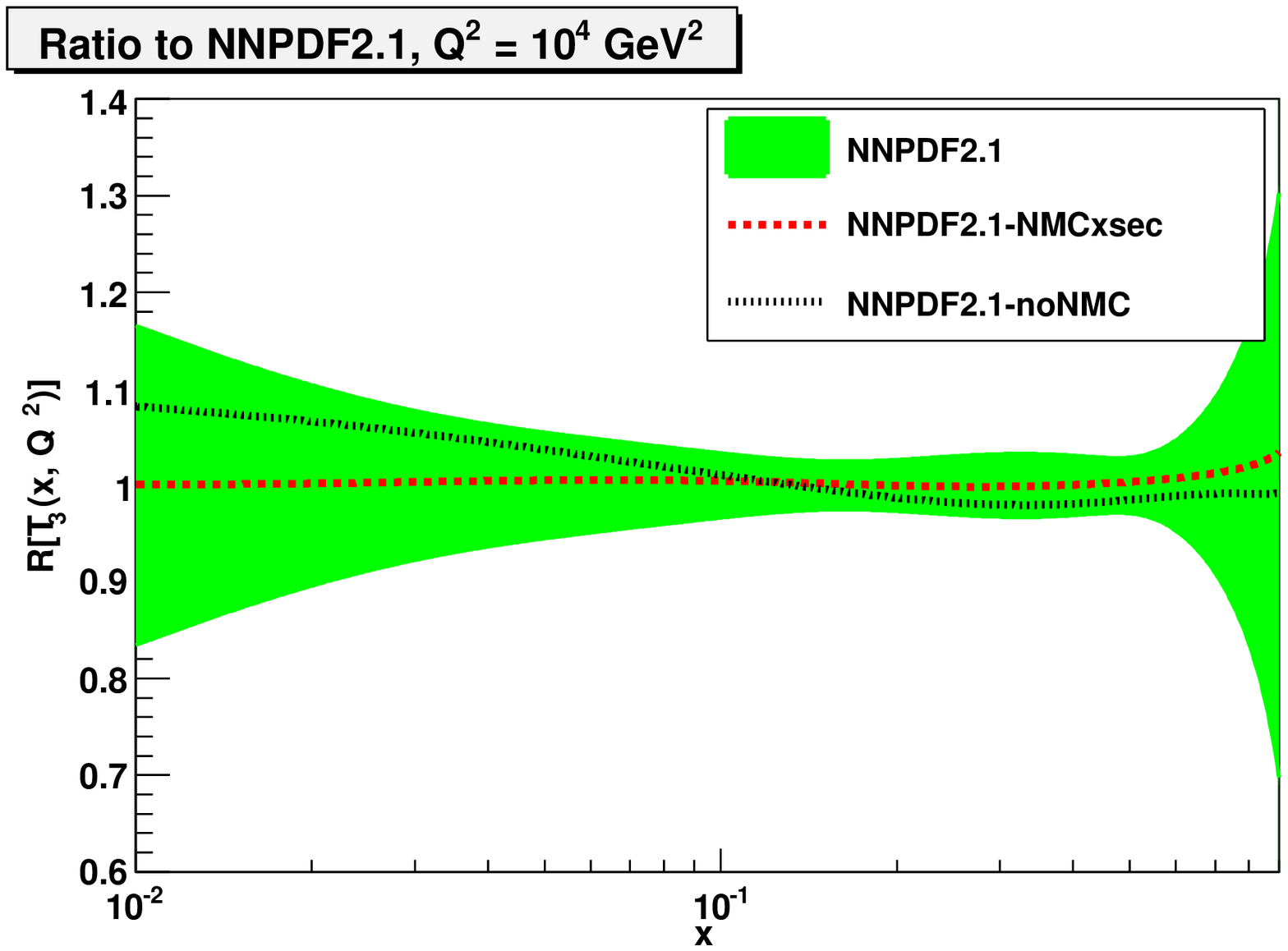,width=0.48\textwidth}
\epsfig{file=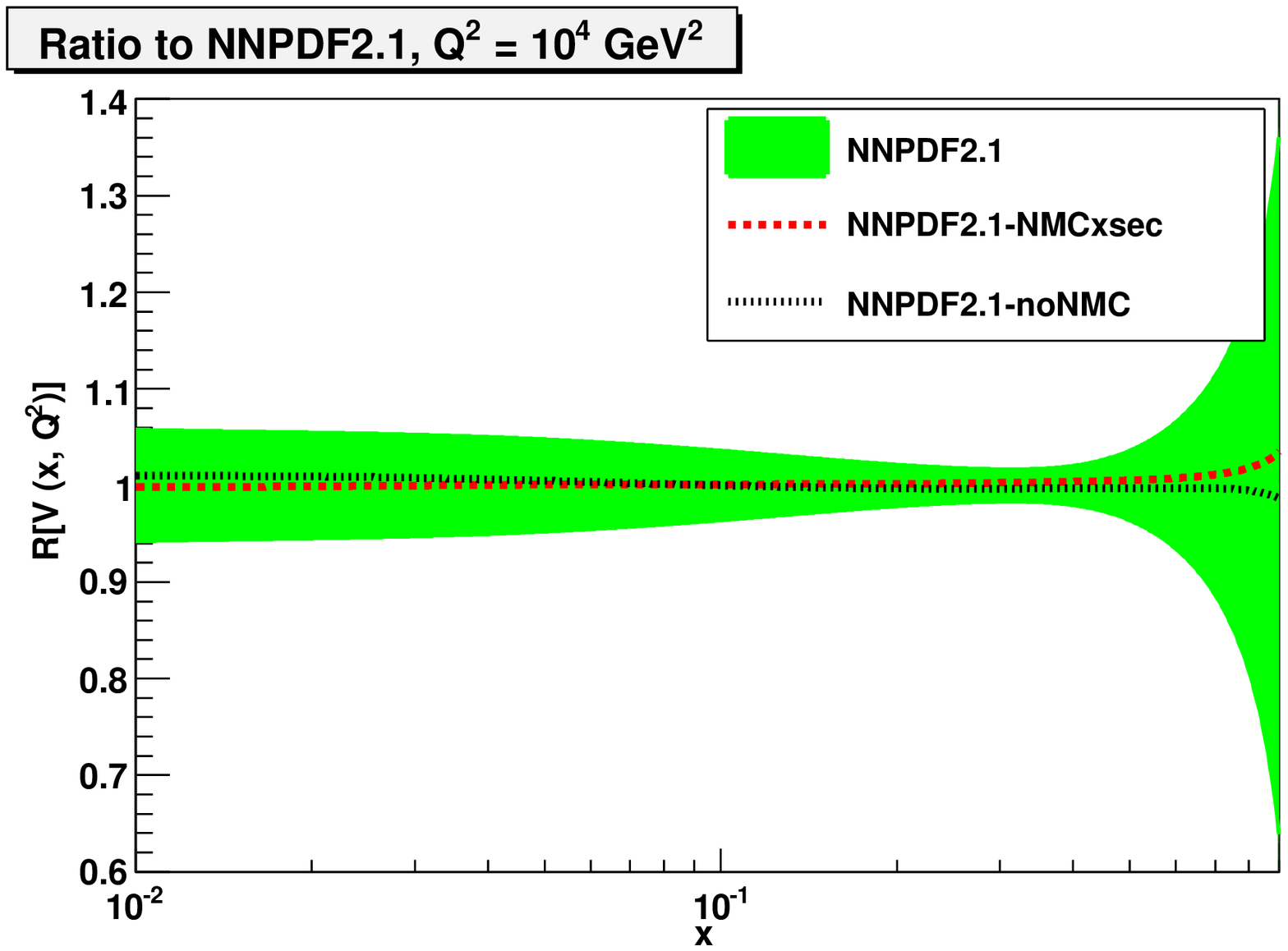,width=0.48\textwidth}
\caption{\small 
NLO PDFs determined using NMC cross-section data (long dashes) and
no NMC data (short dashes) shown as ratio to the default NNPDF2.1 set
(determined using NMC structure function data)
at $Q^2=10^4$ GeV$^2$: 
singlet $\Sigma(x,Q^2)$ (top, left), gluon $g(x,Q^2)$ (top, right), 
triplet $T_3(x,Q^2)$ (bottom, left) and total valence $V(x,Q^2)$
(bottom, right).
\label{fig:pdfplots} }
\end{center}
\end{figure}

\begin{figure}[t]
\begin{center}
\epsfig{file=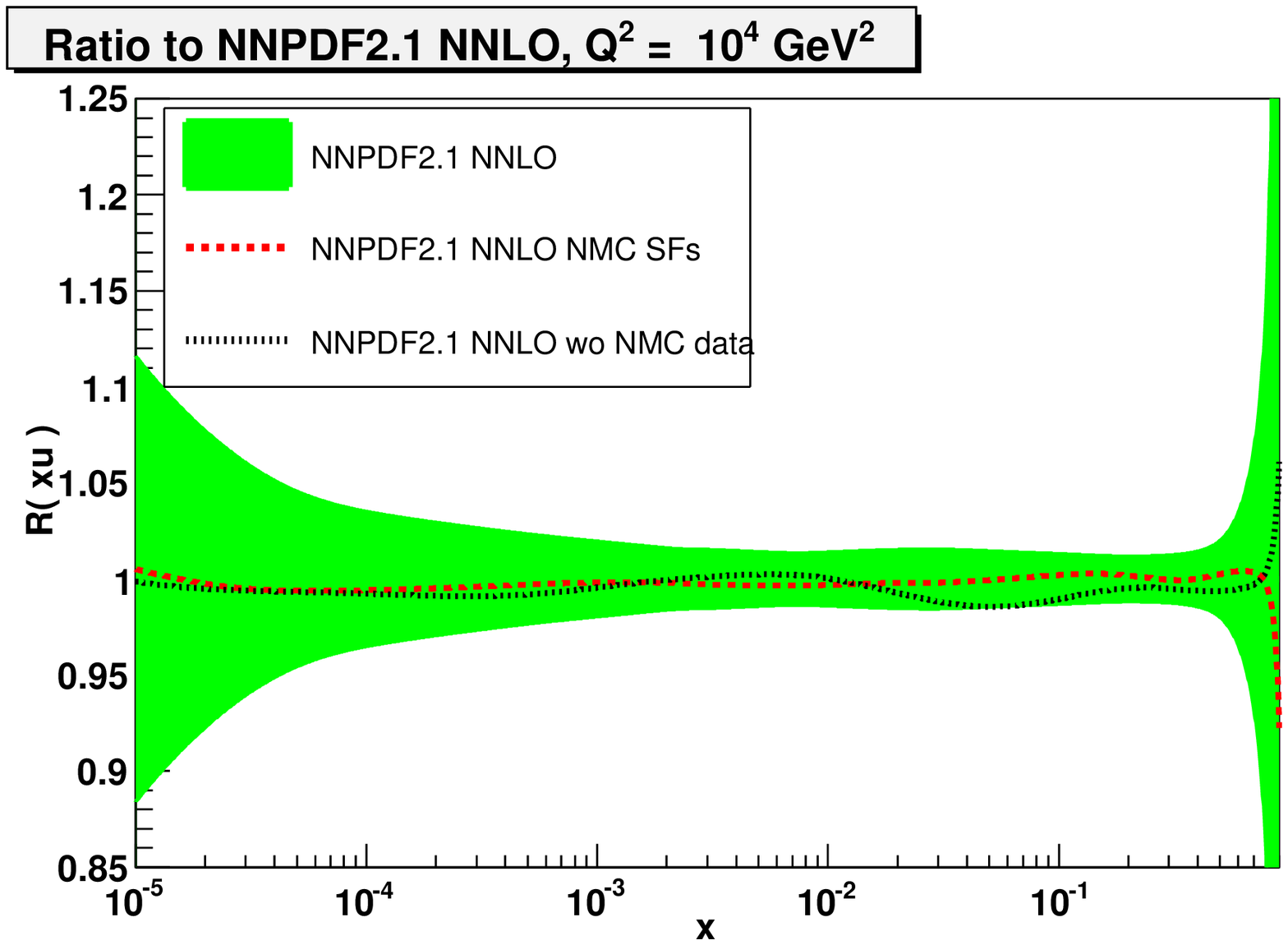,width=0.48\textwidth}
\epsfig{file=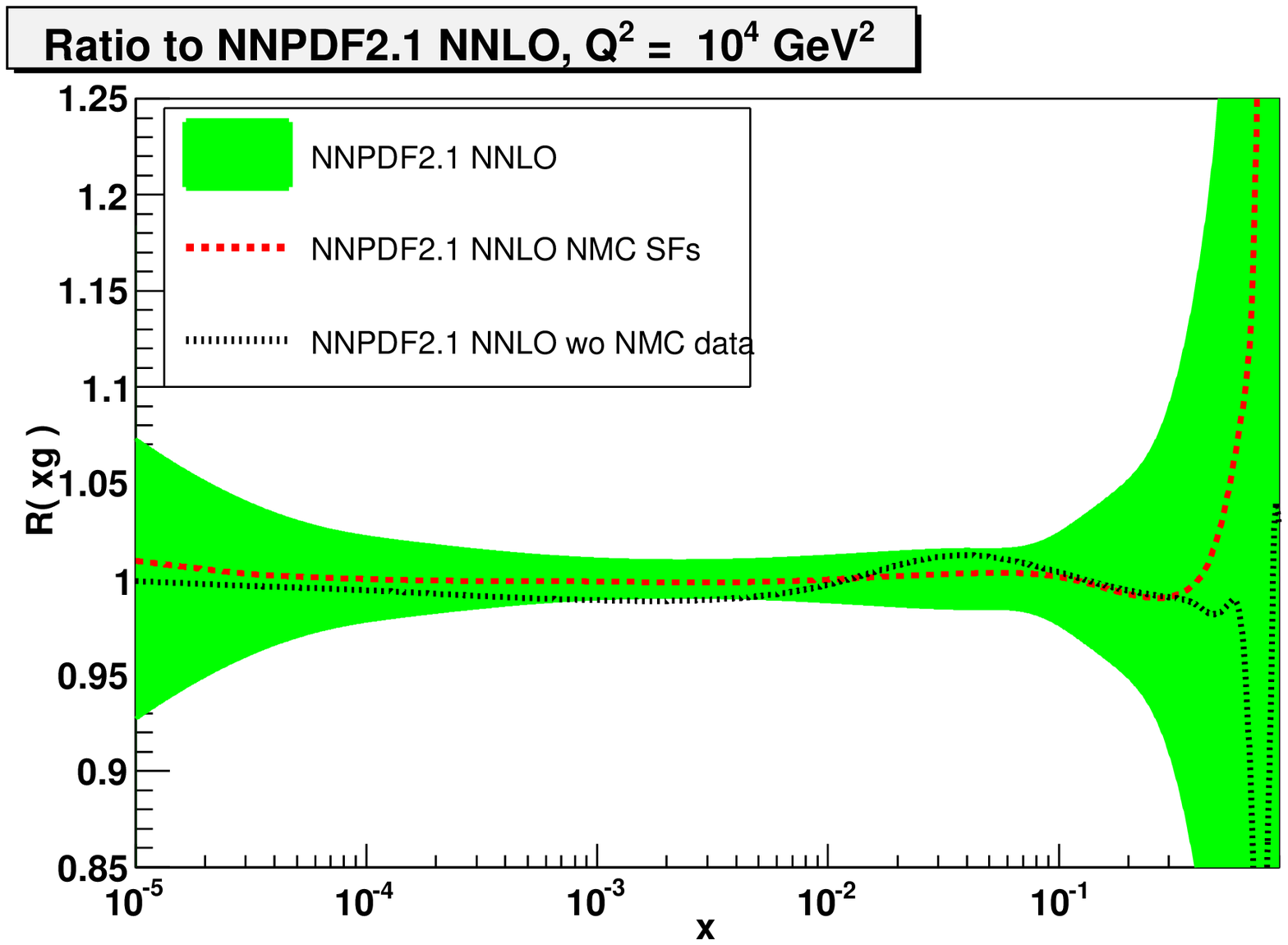,width=0.48\textwidth} \\
\epsfig{file=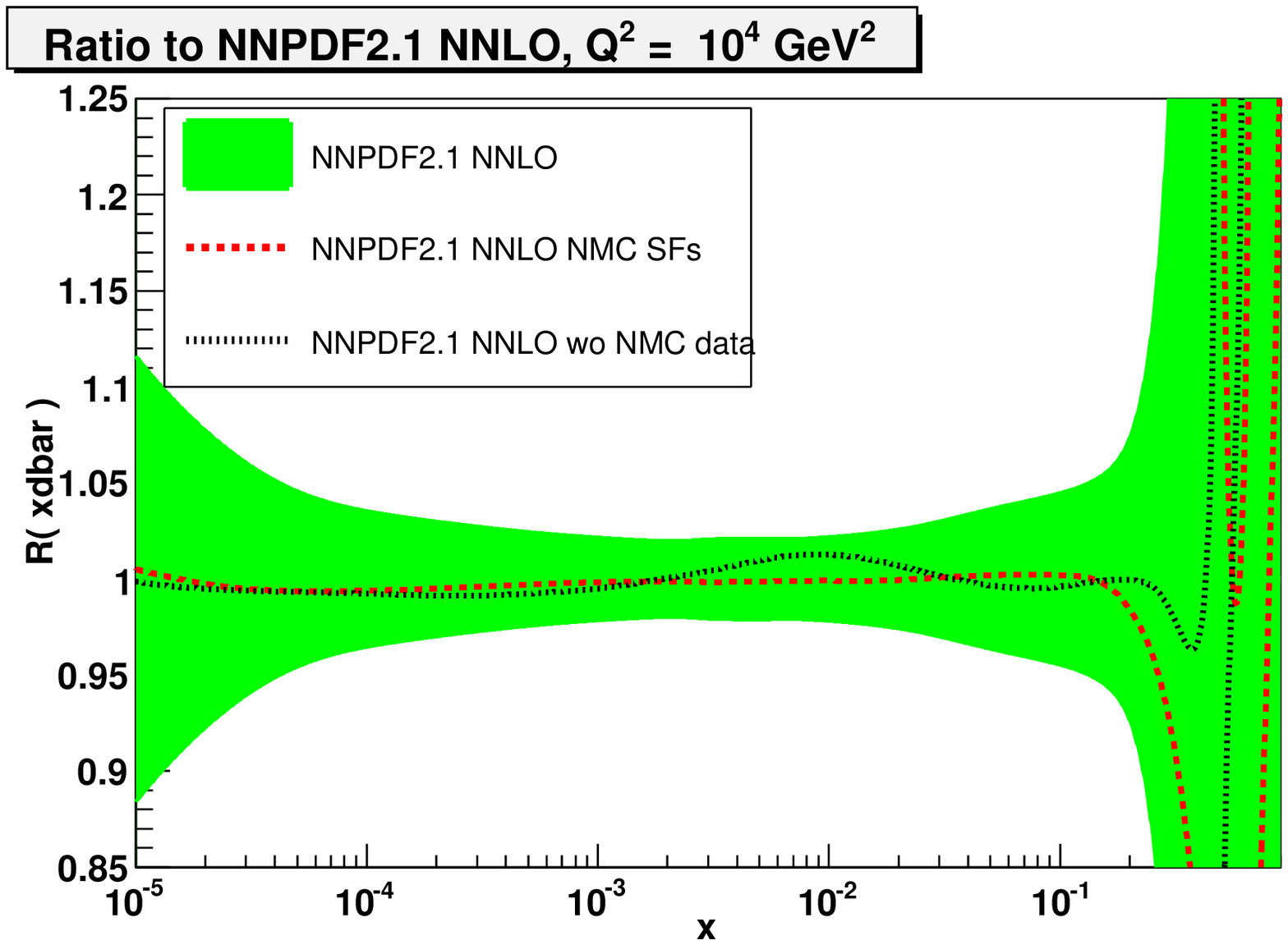,width=0.48\textwidth}
\epsfig{file=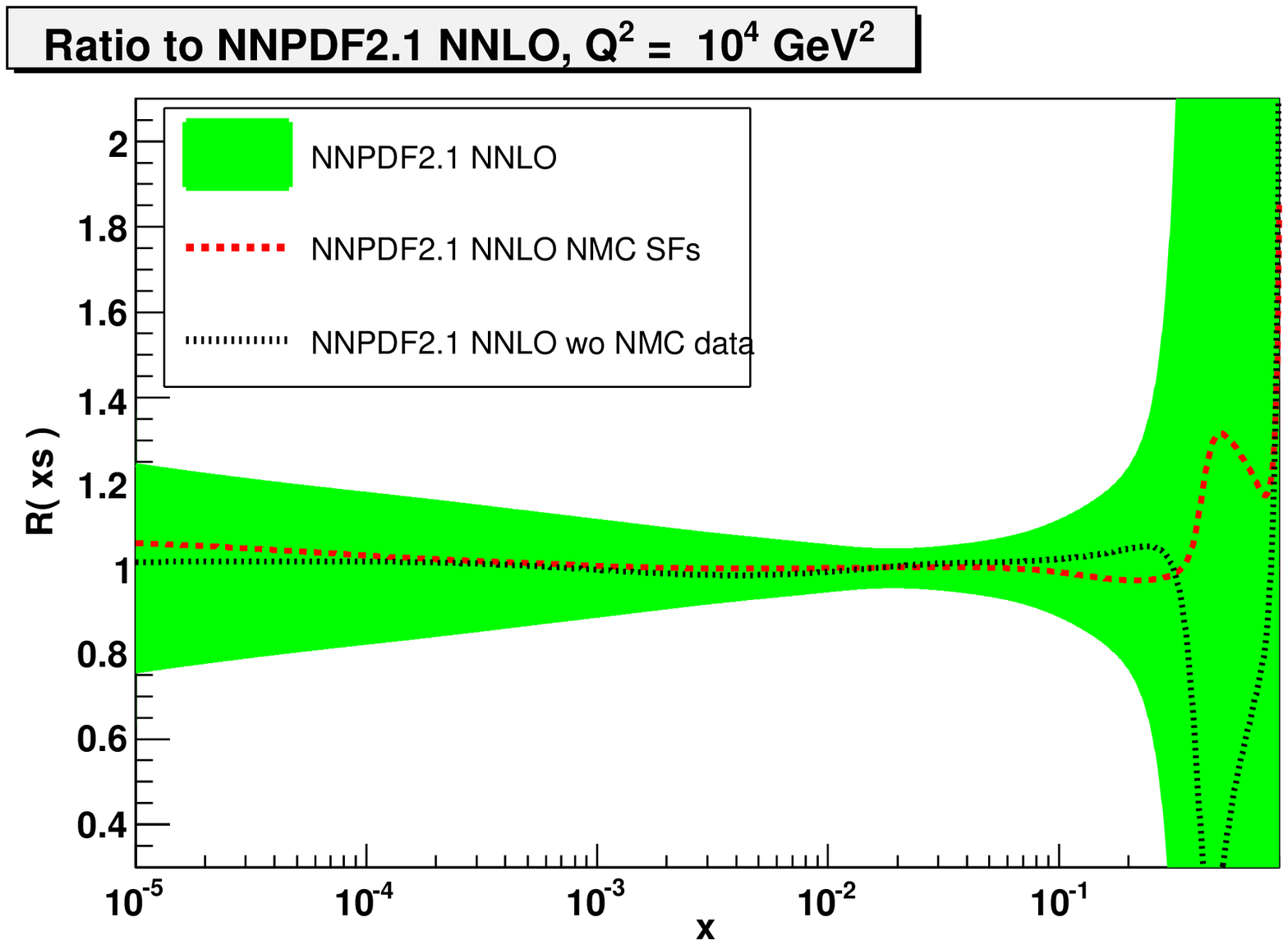,width=0.48\textwidth}
\caption{\small 
NNLO PDFs determined using NMC structure function data (long dashes) and
no NMC data (short dashes) shown as ratio to the default NNPDF2.1 set
(determined using NMC cross-section data)
at $Q^2=10^4$ GeV$^2$: 
up  $u(x,Q^2)$ (top, left), gluon $g(x,Q^2)$ (top, right), 
antdown $\bar{d}(x,Q^2)$ (bottom, left) and strange $s(x,Q^2)$
(bottom, right).
\label{fig:pdfplots-nnlo} }
\end{center}
\end{figure}

Some of these PDFs at NLO and NNLO are compared in
Figs.~\ref{fig:pdfplots} 
 and~\ref{fig:pdfplots-nnlo} respectively, at a typical electroweak
 scale 
$Q^2=10^4~\hbox{GeV}^2$. Differences at higher scale are somewhat
 reduced because of asymptotic freedom, but the general pattern
 observed in the distance plots is clearly reproduced: at NLO
 replacing NMC structure functions with cross-sections has no effect,
 while at NNLO it has an effect which is above the threshold of
 statistical significance, though smaller than the change that would be
 observed if the data changed by an amount compatible with their
 uncertainties. The effect of removing NMC data altogether, both at
 NLO and NNLO, is qualitatively similar but quantitatively somewhat
 larger.

We conclude that at NLO replacing structure functions with cross-sections 
or even removing NMC data altogether has no effect, while at NNLO
replacing structure functions with cross-sections is just above the threshold of
statistical significance, and removing them altogether statistically
significant, though in all cases below the effect of a one sigma
change of the data: this
can be viewed as an upper bound on the possible impact of the treatment of this dataset.

\begin{figure}[t]
\begin{center}
\epsfig{file=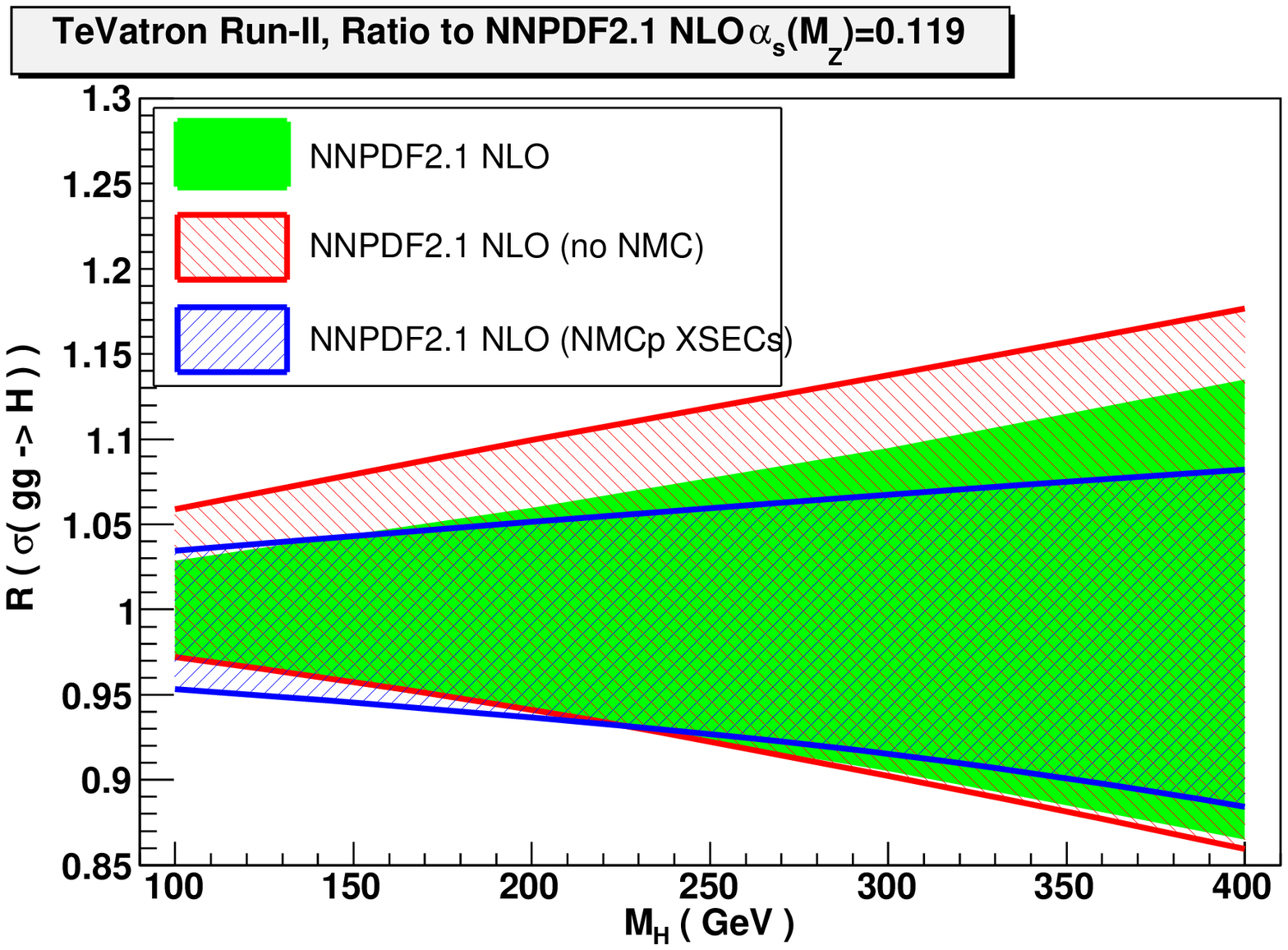,width=0.49\textwidth}
\epsfig{file=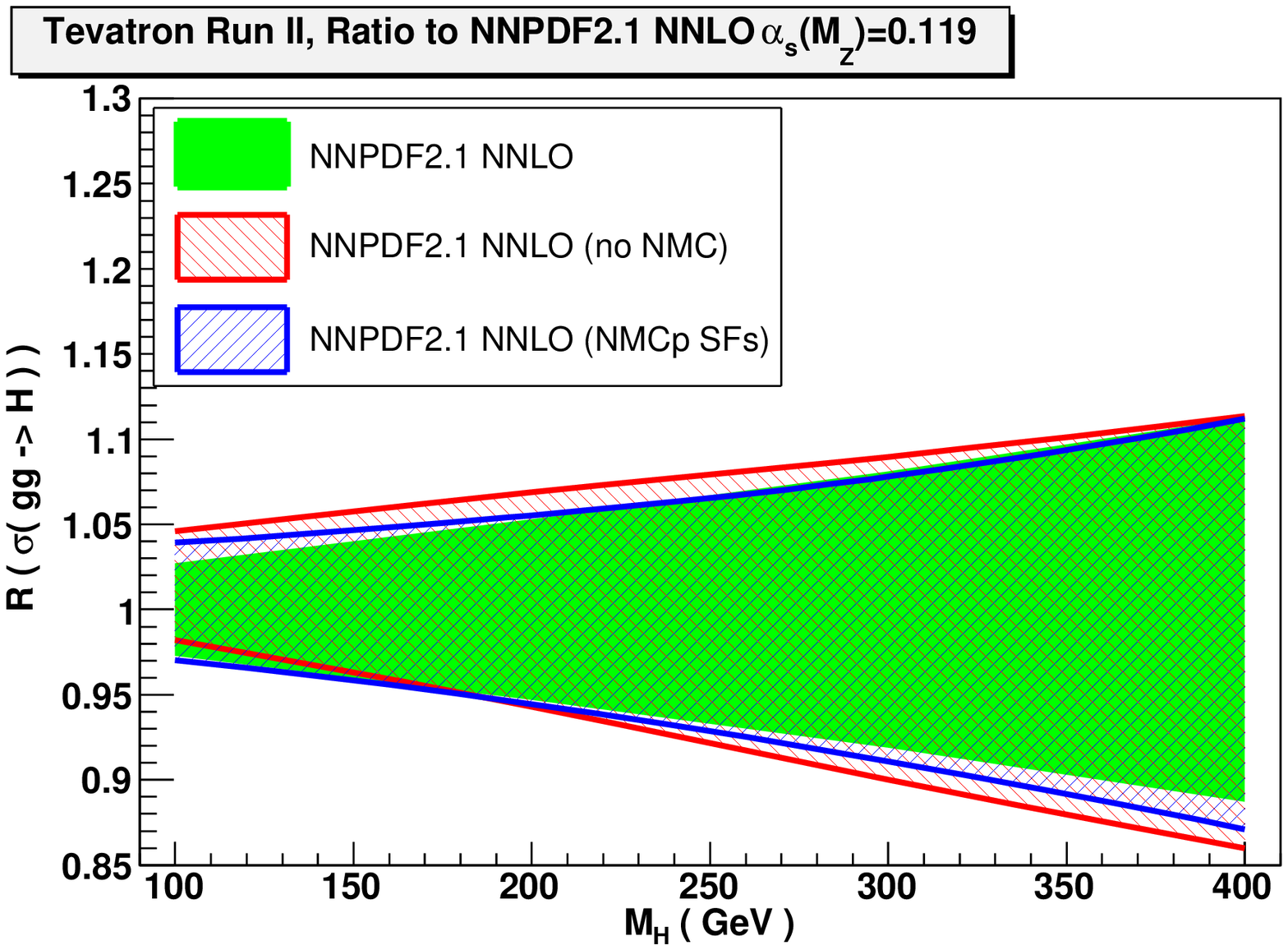,width=0.49\textwidth}
\epsfig{file=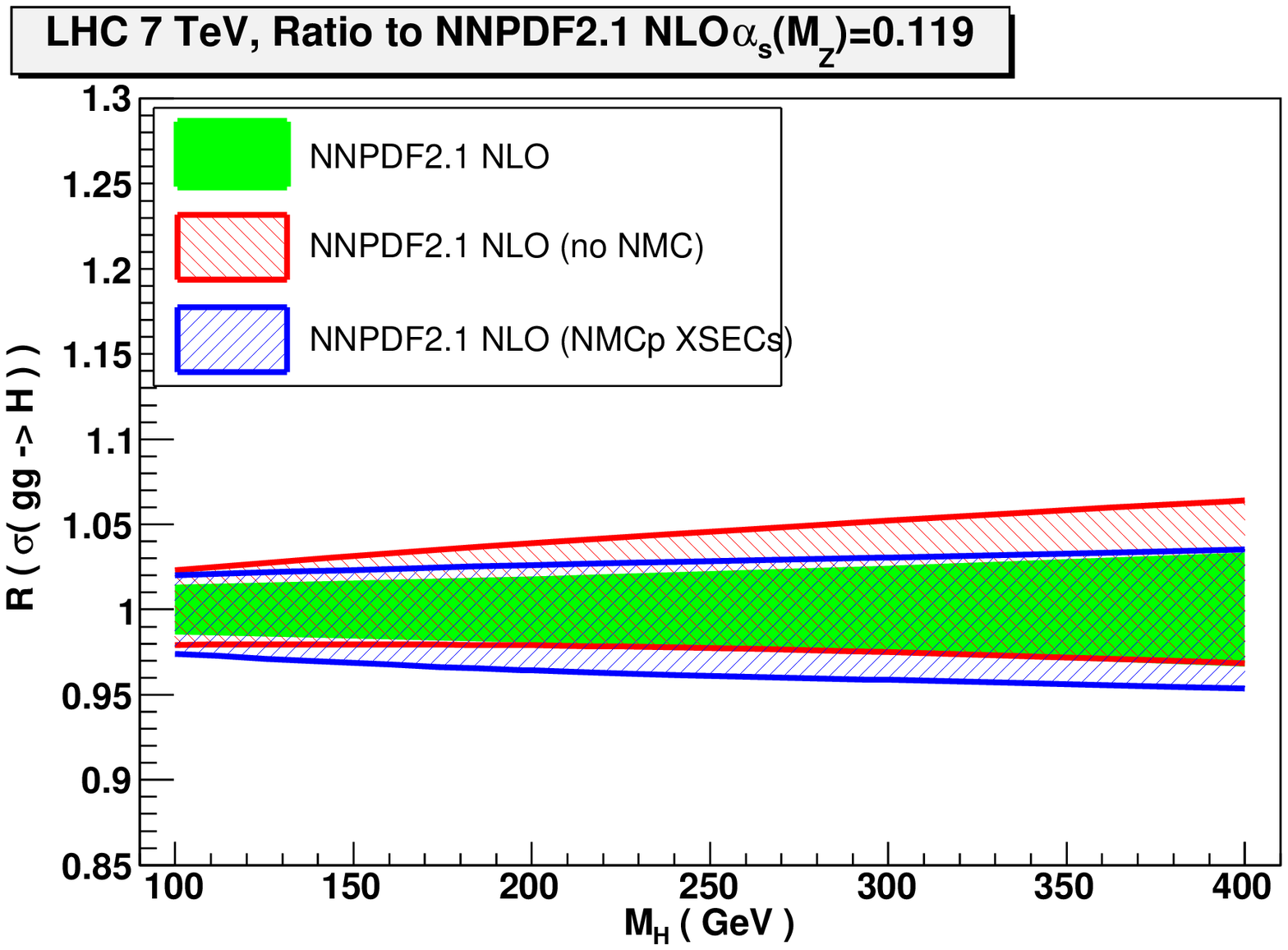,width=0.49\textwidth}
\epsfig{file=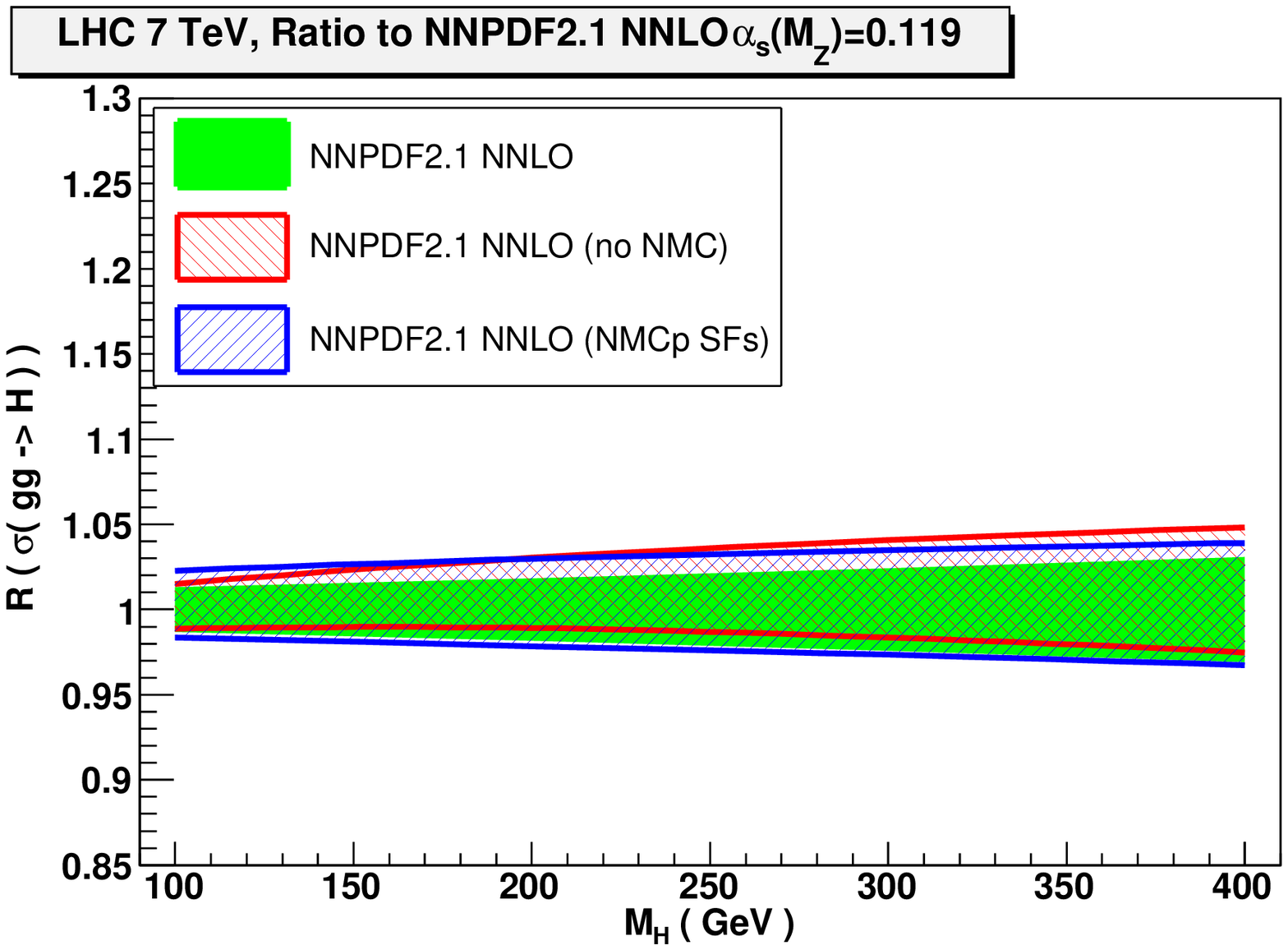,width=0.49\textwidth}
\caption{\small 
The  Higgs cross-section in gluon fusion at the Tevatron
Run II (top) and at the LHC 7 TeV (bottom). Left:
 the reference NNPDF2.1 NLO (NMC structure functions) 
compared to the NLO fits with NMC cross-sections and no NMC
data, shown as a ratio to the reference. Right:   the reference NNPDF2.1N NLO (NMC cross-sections) 
compared to the NLO fits with NMC structure functions and no NMC
data, shown as a ratio to the reference. All uncertainties shown are
one sigma.
\label{fig:higgsplot-nnlo} }
\end{center}
\end{figure}

The main implication of the study of Ref.~\cite{Alekhin:2011ey}, and
the reason for the  
the ensuing debate, was that the Higgs production cross-section via
gluon-gluon fusion may change as a consequence of the treatment of NMC
data by an amount which may invalidate current Higgs exclusion limits.
To verify what happens in our case,
we have recomputed the Higgs production cross-section
using the various PDF sets discussed here, using
the code of
Refs.~\cite{Bonciani:2007ex,Aglietti:2006tp}, for a range
of Higgs masses between 100 and 400 GeV.
We show results for the Tevatron and the LHC 7 TeV
in Fig.~\ref{fig:higgsplot-nnlo}; all uncertainties shown
are 68\% confidence levels. 
We see that the replacement of NMC structure functions by
cross-sections has no impact on the Higgs production
cross-section, and that even removing all NMC data
leads to a shift much smaller than the nominal PDF
uncertainties, with a slight increase of these uncertainties. Again,
this can be viewed as a (conservative) estimate 
of the differences arising from the different treatments of the
NMC data.

Let us finally compare our results with those of
Ref.~\cite{Alekhin:2011ey}. In that reference,
the
value of $\alpha_s$ was determined together with the PDFs, and  the
best-fit $\alpha_s$ was found to change significantly according to the
treatment of the NMC data. In particular, the change of the  best-fit
$\alpha_s$ value was 
found to be of order of 1.5 sigma at NLO and 2.3 sigma at NNLO, with
an increase of
the Higgs cross-section at the LHC by 4\% (i.e. about one sigma)  at NLO and 9\%
(i.e. about 2.7 sigma)  at NNLO when the NMC cross-section data are
replaced by structure function data.
In order to assess quantitatively how much of this
 change in Higgs cross-section is just due to the different value of
 $\alpha_s$ a comparison between ABKM sets with fixed value of
 $\alpha_s$ but different treatment of NMC data would be
 necessary. These sets are at present not available. However, a simple
 estimate (which at NLO is in fact quite
 accurate~\cite{Demartin:2010er}) can be obtained by noting that, 
 based on the size of the NLO and NNLO $K$-factors one expects that a
 percentage change $\Delta\alpha$ in the value of $\alpha_s$, if everything
 else is kept fixed, leads to
 a percentage shift of the Higgs cross-section 
$\Delta\sigma\approx 2.5 \Delta\alpha$ at NLO
 and $\Delta\sigma\approx 2.8 \Delta\alpha$ at NNLO. Based on this,
 one would estimate that about 90\% of the cross-section increase seen  in
Ref.~\cite{Alekhin:2011ey} at
 NLO when structure function data replace
cross-section data and about  80\% of the increase at NNLO, is just 
due to the change in
 value of $\alpha_s$. The residual change, due to the PDFs, is still
 perhaps somewhat larger than that which we observe, but qualitatively
 more in line with it.

We conclude that we do not support the conclusion that the treatment
of NMC data may affect the Higgs cross-section and thus exclusion
limits in any significant way. Of course, if the value of $\alpha_s$
is varied by a very large amount, then the cross-section and ensuing
limits are significantly affected. In this respect, it should be
noticed that  
the best-fit $\alpha_s(M_z)=0.1135$ value of Refs.~\cite{Alekhin:2009ni,Alekhin:2011ey} at NNLO differs by 
more than 7 sigma from the PDG value $\alpha_s(M_z)=0.1184$~\cite{Bethke:2009jm}, in units of the latter's uncertainty $\Delta
\alpha_s=0.0007$.  The Higgs working group~\cite{Dittmaier:2011ti}
following PDF4LHC~\cite{Botje:2011sn}, recommends to use a more conservative $\Delta
\alpha_s=0.0012$, but even so the value of
Refs.~\cite{Alekhin:2009ni,Alekhin:2011ey}  differs by more than four
sigma from the PDG average.
A NLO determination of $\alpha_s$ based on the
NNPDF2.1 PDF fit~\cite{Lionetti:2011pw} 
leads to values of $\alpha_s$ which are in good agreement with
the PDG value, both when  the global dataset ($\alpha_s(M_z)=0.1191$)
and 
deep-inelastic data only ($\alpha_s(M_z)=0.1177$) are used. Given
that, as we have just shown,  the treatment of NMC data has
no statistically significant impact on the NLO analysis, it is
exceedingly unlikely that the NNPDF NLO determination of $\alpha_s$ might
depend on how the NMC data are treated.
It will be
interesting to repeat the analysis of Ref.~\cite{Lionetti:2011pw} at
NNLO. The  stability of NNPDF2.1 PDFs when going from NLO to
NNLO~\cite{Ball:2011uy} suggests that results should not change
dramatically. 


In summary, we find that the effect of the  treatment of NMC data on
NNPDF2.1 PDFs is of no
statistical significance at NLO, and just about statistically
significant at NNLO though by at least a factor three smaller than the
typical PDF uncertainty due to propagated data uncertainties.
The effect on the Higgs
production cross-section is accordingly negligible. 
Even removing NMC data altogether 
has a moderate effect on NNPDF2.1 PDFs, which even at NNLO remains 
below one sigma.  
The considerable stability of the NNPDF2.1
results is due both to the use of a very wide dataset which includes
DIS, Drell-Yan, weak vector boson production and inclusive jet data, which reduces the
dependence of our results on any particular dataset, and to the extremely 
flexible neural network parametrization which eliminates the parametrization bias which 
might otherwise lead to instabilities on small shifts in input data.

\bigskip

{\bf\noindent  Acknowledgments \\} We thank S.~Alekhin for
correspondence and discussions on the matter of this paper.
M.U. is supported by the Bundesministerium f\"ur Bildung and Forschung (BmBF) of the Federal 
Republic of Germany (project code 05H09PAE).
This work was partly supported by the Spanish MEC FIS2007-60350 grant. 
We would like to acknowledge the use of the computing resources provided 
by the Black Forest Grid Initiative in Freiburg.


\begin{thebibliography}{100}

\bibitem{Alekhin:2011ey}
S. Alekhin, J. Blumlein and S. Moch,
\newblock (2011), arXiv:1101.5261.

\bibitem{Arneodo:1996qe}
New Muon Collaboration, M. Arneodo et~al.,
\newblock Nucl. Phys. B483 (1997) 3, hep-ph/9610231.

\bibitem{Arneodo:1996kd}
New Muon Collaboration, M. Arneodo et~al.,
\newblock Nucl. Phys. B487 (1997) 3, hep-ex/9611022.

\bibitem{Baglio:2011hc}
  J.~Baglio, A.~Djouadi and R.~Godbole,
\newblock (2011),  arXiv:1107.0281.

\bibitem{Thorne:2011kq}
  R.~S.~Thorne and G.~Watt,
\newblock (2011),  arXiv:1106.5789.


\bibitem{Ball:2011mu}
The NNPDF Collaboration, R.D. Ball et~al.,
\newblock Nucl. Phys. B849 (2011) 296, arXiv:1101.1300.

\bibitem{Ball:2011uy}
  R.~D.~Ball {\it et al.}  [The NNPDF Collaboration],
\newblock (2011)  arXiv:1107.2652.


\bibitem{Alekhin:2009ni}
S. Alekhin et~al.,
\newblock Phys. Rev. D81 (2010) 014032, arXiv:0908.2766.

\bibitem{Forte:2010dt}
S. Forte,
\newblock  Acta Phys. Polon. B41 (2010) 2859, arXiv:1011.5247.

\bibitem{HustonPDF4LHC}
J.~Huston, talk at the PDF4LHC workshop, CERN, March 2011
{\tt\footnotesize http://indico.cern.ch/getFile.py/access?contribId=2\&resId=1\&materialId=slides\&confId=127425}

\bibitem{Whitlow:1991uw}
L.W. Whitlow et~al.,
\newblock Phys. Lett. B282 (1992) 475.

\bibitem{Whitlow:1990gk}
L.W. Whitlow et~al, 
\newblock Phys. Lett. 250B (1990) 193.


\bibitem{Bethke:2009jm}
S. Bethke,
\newblock Eur. Phys. J. C64 (2009) 689, arXiv:0908.1135.

\bibitem{Demartin:2010er}
F. Demartin et~al.,
\newblock Phys. Rev. D82 (2010) 014002, arXiv:1004.0962.

\bibitem{Ball:2010de}
{The NNPDF collaboration}, R.D. Ball et~al.,
\newblock Nucl. Phys. B838 (2010) 136, 1002.4407.

\bibitem{Alekhin:2011sk}
S. Alekhin et~al.,
\newblock (2011), arXiv:1101.0536.

\bibitem{Dittmaier:2011ti}
LHC Higgs Cross Section Working Group, S. Dittmaier et~al.,
\newblock (2011), arXiv:1101.0593.


\bibitem{DelDebbio:2007ee}
The NNPDF collaboration, L. Del~Debbio et~al.,
\newblock JHEP 03 (2007) 039, hep-ph/0701127.

\bibitem{Ball:2008by}
The NNPDF Collaboration, R.D. Ball et~al.,
\newblock Nucl. Phys. B809 (2009) 1, arXiv:0808.1231.

\bibitem{Ball:2009mk}
The NNPDF collaboration, R.D. Ball et~al.,
\newblock Nucl. Phys. B823 (2009) 195, arXiv:0906.1958.




\bibitem{Ball:2009qv}
The NNPDF collaboration, R.D. Ball et~al.,
\newblock JHEP 05 (2010) 075, arXiv:0912.2276.

\bibitem{Forte:2002fg}
S. Forte et~al.,
\newblock JHEP 05 (2002) 062, hep-ph/0204232.


\bibitem{Bonciani:2007ex}
R. Bonciani, G. Degrassi and A. Vicini,
\newblock JHEP 11 (2007) 095, arXiv:0709.4227.

\bibitem{Aglietti:2006tp}
U. Aglietti et~al.,
\newblock JHEP 01 (2007) 021, hep-ph/0611266.

\bibitem{Botje:2011sn}
  M.~Botje {\it et al.},
\newblock  (2011) 1101.0538.

\bibitem{Lionetti:2011pw}
S. Lionetti et~al.,
\newblock Phys. Lett. B701 (2011) 346, arXiv:1103.2369.

\end{thebibliography}
\end{document}